\documentclass[twocolumn]{jpsj2} 
\title{Universal temperature dependence of electron number in one-dimensional Hubbard model}

\author{Jun \textsc{Maeda}$^{1}$\thanks{Present address: Fujitsu Limited, Kawasaki, Kanagawa 211-8588} and 
Sei-ichiro \textsc{Suga}$^{1,2}$ }

\inst{$^{1}$Department of Applied Physics, Osaka University, Suita, Osaka 565-0871 \\
$^{2}$Department of Materials Science and Chemistry, University of Hyogo, Himeji, Hyogo 671-2280 \\
}

\abst{
We investigate the temperature region in which a Tomonaga-Luttinger liquid (TLL) description of the charge sector of the one-dimensional Hubbard model is valid. By using the thermodynamic Bethe ansatz method, electron number is calculated at finite temperatures and fixed chemical potential. 
We observe maximum electron number as a function of temperature close to the chemical potential of the upper critical value that corresponds to half filling. As the chemical potential approaches the upper critical value from below, the temperature $(T_{\rm M})$ at which the electron number shows its maximum asymptotically approaches a universal relation. 
We show that, below the energy corresponding to $T_{\rm M}$, the charge excitation spectrum nearly obeys a linear dispersion relation. The results demonstrate that $T_{\rm M}$ marks the important temperature below which TLL is realized.  
}
\kword{1D Hubbard model, electron number, excitation spectra, Tomonaga-Luttinger liquid, Bethe ansatz}

\begin{document}
\maketitle

\section{\label{sec:level1}Introduction}
One-dimensional (1D) gapped many-body systems have attracted much attention. Interesting aspects have been revealed by the collaboration of theorists and experimentalists. 
For 1D gapped spin systems, it was argued theoretically that the characteristic features of each system appear  conspicuously in critical and dynamical properties above the critical magnetic field $(H_{\rm c})$ where the gap is closed \cite{chitra}. Above the critical field, low-energy properties can be described as a Tomonaga-Luttinger liquid (TLL). The characteristic behavior depending on the model can be observed in, for example, the divergence of NMR relaxation rate with decreasing temperature \cite{chitra,haga}. 
In fact, for the $S=1$ Haldane-gap material ${\rm (CH_3)_4NNi(NO_2)_3}$ \cite{goto1,goto2} and the $S=1/2$ bond-alternating chain material pentafluorophenyl nitronyl nitroxide \cite{izumi} a divergence of NMR relaxation rate with decreasing temperature was observed. The field dependence of the divergence exponent was discussed in comparison with theoretical results \cite{sakai,suzuki1,suzuki2}. 
Recent experimental advances have further allowed the observation of typical TLL behavior of thermodynamic quantities in 1D gapped spin systems. The specific heat of the quasi-1D $S=1$ bond-alternating antiferromagnet Ni(C$_9$H$_{24}$N$_4$)(NO$_2$)ClO$_4$ was measured at magnetic fields above $H_{\rm c}$ \cite{hagiwara}. Low-temperature specific heat was found to be proportional to temperature, indicative of the gapless dispersion relation. The field dependence of its coefficient was in good agreement with the numerical result based on conformal field theory (CFT) \cite{hagiwara}.

Although the characteristic behavior of TLL has been studied intensively, it has been difficult to determine the temperature region where the TLL picture is valid from the gapped spin Hamiltonian itself. To develop precise comparison between the theoretical and experimental results for the TLL, this issue has to be resolved. 
In a recent study, it was shown that the magnetization minimum appears as a function of temperature close to $H=H_{\rm c}$ in 1D gapped spin systems with axial symmetry \cite{maeda}. The magnetization minimum marks the important temperature below which TLL is valid. It was demonstrated further that the temperature for the magnetization minimum approaches a universal relation as $H$ approaches $H_{\rm c}$ from above. The arguments shed light on determining the temperature region of the TLL by theoretical and experimental methods.
Actually, the predicted features were confirmed theoretically for the $S=1$ Haldane-gap system \cite{maeda} and the 1D $S=1/2$ Heisenberg model with a large Ising anisotropy \cite{suga} in magnetic fields.

The origin of the magnetization minimum and universal relation was elucidated using the effective theory near the critical field and the low-temperature expansion of the free energy based on CFT \cite{maeda}. Since they are based on the fundamental nature of the TLL, the analyses seem to be applicable to general 1D gapped systems with U(1) symmetry close to the critical point.  
To confirm this consideration, in this paper, we investigate the electron number of the 1D Hubbard model at finite temperatures using the Bethe ansatz method. The charge sector of the 1D Hubbard model is described by $c=1$ CFT with U(1) symmetry \cite{kawakami,frahm}. Here, $c$ is the central charge. 
In $\S$2, we outline the model and the method for the calculation. 
In $\S$3, we present the results. 
It is shown that a maximum appears in the temperature dependence of the electron number close to half filling. The temperature $(T_{\rm M})$ for the maximum electron number approaches the universal relation, as the system approaches half filling. We find that, below the excitation energy corresponding to $T_{\rm M}$, the linear dispersion relation of the charge excitation spectrum is nearly satisfied.  
Recently, various 1D organic conductors have been synthesized. For such compounds, pressure-induced quantum phase transitions from a charge-gapped state to the TLL state have been investigated experimentally \cite{org}. The carrier number close to the critical point is expected to show a minimum/maximum as a function of temperature and the temperature for the minimum/maximum carrier number plays a role in determining the temperature region of the TLL. 
Section 4 provides a summary.

\section{\label{sec:level2}Model and Method}
Let us consider the 1D Hubbard model 
\begin{eqnarray}
\mathcal{H} = -t\sum_{\langle i,j \rangle, \sigma} c^{\dagger}_{i \sigma}c^{}_{j \sigma} +4U\sum_{i}n_{i \uparrow}n_{i \downarrow} -\mu\sum_{i, \sigma}n_{i, \sigma},
\label{Ham}
\end{eqnarray}
where the subscript $\langle i,j \rangle$ indicates the sum over nearest-neighbor sites, $c^{}_{i \sigma}$ annihilates an electron with spin $\sigma(=\uparrow, \downarrow)$ at the $i$th lattice site, $n_{i, \sigma}=c^{\dagger}_{i \sigma}c^{}_{i \sigma}$, and $\mu$ is the chemical potential. The hopping integral $t$ is given in units of energy. 
As $\mu$ increases between its lower and upper critical values ($\mu_{\rm c1}=-2.0<\mu<\mu_{\rm c2}$), electron number ($n$) increases monotonically from zero to unity. At half filling ($n=1$), the Mott insulating state appears. 
The upper critical value of the chemical potential is obtained by the Bethe ansatz solution at zero temperature $(T=0)$ as 
$\mu_{\rm c2} =F(Q)/D(Q)|_{Q=\pi}$, where $F(k)$ and $D(k)$ are obtained using the integral equations 
\begin{eqnarray}
F(k) = &-& 2\cos k  \nonumber \\
       &+& \int_{-Q}^{Q} dk^{\prime} 
              \cos k^{\prime} R(\sin k-\sin k^{\prime}) F(k^{\prime}), 
\label{eqn:de}
\end{eqnarray}
\begin{eqnarray}
D(k) = 1 + \int_{-Q}^{Q} dk^{\prime} 
             \cos k^{\prime} R(\sin k-\sin k^{\prime}) D(k^{\prime}), 
\label{eqn:rho}
\end{eqnarray}
with $k$ being the charge rapidity and $R(x)=(1/2\pi)\int_{-Q}^{Q} d\omega e^{-i\omega x} \left(1+e^{2U |\omega|} \right)^{-1}$. 
The cutoff $Q$ is obtained under the condition $\kappa(Q)=0$ for the dressed energy of charge rapidity. The dressed energy $\kappa(k)$ is obtained using the integral equation, 
\begin{eqnarray}
\kappa(k) = &-& 2\cos k -\mu
          - \int_{-\infty}^{\infty} d\lambda a_1(\lambda-\sin k) \nonumber \\
           &\times& \int_{-Q}^{Q} dk^{\prime}\cos k^{\prime} 
                   S(\lambda-\sin k^{\prime}) \kappa (k^{\prime}), 
\label{eqn:de}
\end{eqnarray}
where $\lambda$ is the spin rapidity, $a_1(\lambda)=(U/\pi)/(\lambda^2+U^2)$, and $S(x)=(1/4U){\rm sech} (\pi \lambda/2U)$.

Thermodynamic quantities of the 1D Hubbard model are formulated using the thermodynamic Bethe ansatz method \cite{takahashi}. 
In the thermodynamic limit, an infinite set of nonlinearly coupled integral equations is derived for three types of pseudoenergies. Two of them are for two types of spin rapidities and the third one is for charge rapidity. 
The free-energy functional is expressed in terms of pseudoenergies. 
To perform the numerical calculation, we necessarily cut off two infinite sets of coupled integral equations at finite numbers, which are set at 60 and 5 \cite{comm}. We employ the  numerical technique developed by Usuki {\it et al.} \cite{usuki}, where thermodynamic quantities are accurately calculated without numerical differentiation of the free-energy functional. 
The range of spin rapidity is set to be $[-80, 80]$, which is divided into 1600 points. The range of charge rapidity $[-\pi, \pi]$ is divided into 800 points. 
The iterative calculations for pseudoenergies are carried out until they all converge within a relative error of $10^{-6}$. Their derivatives with respect to the chemical potential are then calculated in the same way within a relative error of $10^{-4}$.

Using the thus-obtained pseudoenergies and their derivatives, we calculate electron number at finite temperatures. In the following calculations, we set $U=2.0$, which leads to $\mu_{\rm c2}=1.66$.

\section{Results and Discussion}
\begin{figure}[htb]
\begin{center}
\includegraphics[clip, width=7.5cm]{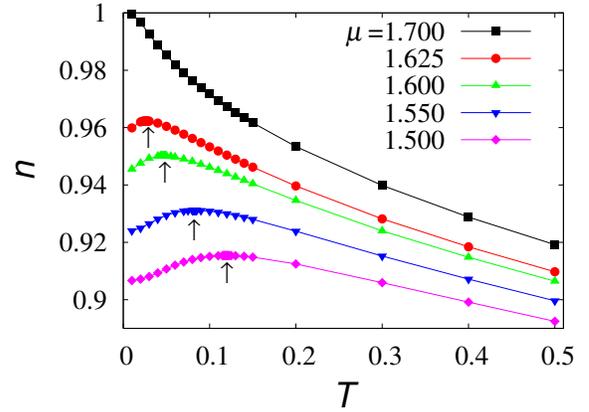}
\caption{(Color online) Temperature dependence of electron number for a given chemical potential close to and above the upper critical value $(\mu_{\rm c2}=1.66)$. 
Electron number takes a maximum at the temperature indicated by an arrow.}
\label{mg}
\end{center}
\end{figure}
\begin{figure}[htb]
\begin{center}
\includegraphics[clip, width=7.5cm]{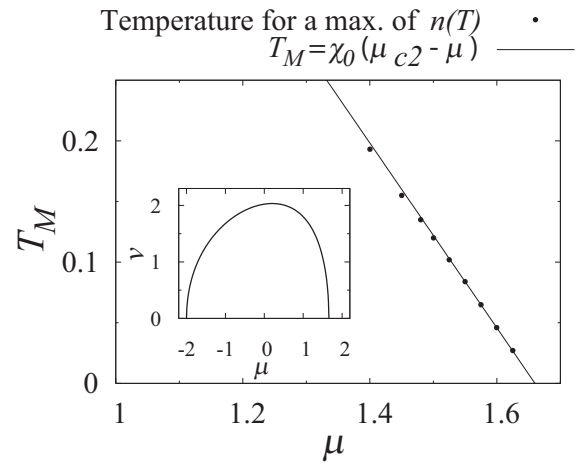}
\caption{Temperature for maximum electron number as a function of chemical potential. The solid circles represent $T_{\rm M}$ and the solid line expresses the universal relation $T_{\rm M}=-x_0(\mu-\mu_{\rm c2})$, ($x_0 \sim 0.76238$). 
Inset: velocity as a function of chemical potential. $\mu_{\rm c1}=-2.0$ and $\mu_{\rm c2}=1.66$.}
\label{tm}
\end{center}
\end{figure}
In Fig. \ref{mg}, we show the temperature dependence of electron number for several chemical potentials close to $\mu=\mu_{\rm c2}$. 
For the gapless regime $\mu \lesssim \mu_{\rm c2}$, the electron number shows a nontrivial maximum at low temperatures. As chemical potential approaches the critical value $\mu_{\rm c2}$, $T_{\rm M}$ significantly decreases towards zero.  For the Mott insulating region $\mu \gtrsim \mu_{\rm c2}$, by contrast, electron number shows no maximum and decreases monotonically with increasing temperature. From the low-temperature expansion for the free-energy functional based on CFT \cite{blote,affleck}, electron number takes the form 
\begin{eqnarray}
n=n_0-\frac{\pi}{6v_{\rm c}^2}\frac{\partial v_{\rm c}}{\partial \mu}T^2 + O(T^3), 
\label{eqn:n}
\end{eqnarray}
where $n_0$ is the electron number at $T=0$ and $v_{\rm c}$ is the velocity of the charge excitation. The second term is the leading finite temperature correction. 
Accordingly, the behavior of the electron number near $T=0$ is determined by the sign of $\partial v_{\rm c}/\partial \mu$. 
The velocity for the charge excitation is obtained as 
$v_{\rm c}=[d \kappa(k)/d k]/[2\pi \rho(k)]|_{k=Q}$, where $\rho(k)$ is the distribution function of the charge rapidity obtained using the integral equation,
\begin{eqnarray}
\rho(\lambda) = \frac{1}{2\pi} &+& \cos k \int_{-\infty}^{\infty} d\lambda 
                a_1(\lambda-\sin k) \nonumber \\
                &\times& \int_{-Q}^{Q} dk^{\prime} 
                         S(\lambda-\sin k^{\prime}) \rho(k^{\prime}). 
\label{eqn:rho}
\end{eqnarray}
As shown in the inset of Fig. \ref{tm}, velocity decreases with increasing chemical potential close to $\mu=\mu_{\rm c2}$, leading to a negative $\partial v_{\rm c}/\partial \mu$. As temperature is increased, therefore, the electron number close to $\mu=\mu_{\rm c2}$ increases near $T=0$ and then decreases, yielding a maximum structure.

In Fig. \ref{tm}, $T_{\rm M}$ is shown as a function of chemical potential. As chemical potential approaches $\mu_{\rm c2}$, $T_{\rm M}$ asymptotically approaches the universal relation $T_{\rm M}=-x_0(\mu-\mu_{\rm c2})$, ($x_0 \sim 0.76238$). 
This relation can be derived similarly to that developed for 1D gapped spin systems \cite{maeda}. 
Close to the upper critical value, the low-energy state can be approximated by 
$E(p) \sim -p^2/2m +\mu_{\rm c2}-\mu$, where $m$ is the effective mass of the electron. The electron number for $\mu \lesssim \mu_{\rm c2}$ is obtained using $n={\rm const}.+\sqrt{mT/2\pi}{\rm Li}_{1/2}(-e^{(\mu_{\rm c2}-\mu)/T})$ with ${\rm Li}_{n}(x)$ being the polylogarithm function. 
The differentiation of $n$ with respect to $T$ yields the condition for a maximum electron number, 
$2(\mu_{\rm c2}-\mu)/T_{\rm M}={\rm Li}_{1/2}(-e^{(\mu_{\rm c2}-\mu)/T_{\rm M}})/{\rm Li}_{-1/2}(-e^{(\mu_{\rm c2}-\mu)/T_{\rm M}})$, which leads to the universal relation mentioned above. 
If we replace $\mu$ with the magnetic field, the same universal relation as that derived for the 1D gapped spin system \cite{maeda} except for the sign can be obtained. The sign of the universal relation depends on which critical point, the upper or lower, is paid attention to. 
The magnetization of 1D spin systems can be regarded as the particle number in fermionic particle language. Therefore, the thermodynamic quantity corresponding to the particle number shows a minimum/maximum as a function of the temperature close to the critical point of 1D gapped many-body systems. The temperature for a minimum/maximum approaches the universal relation, as the system approaches the critical point.

\begin{figure}[htb]
\begin{center}
\includegraphics[width=7.5cm]{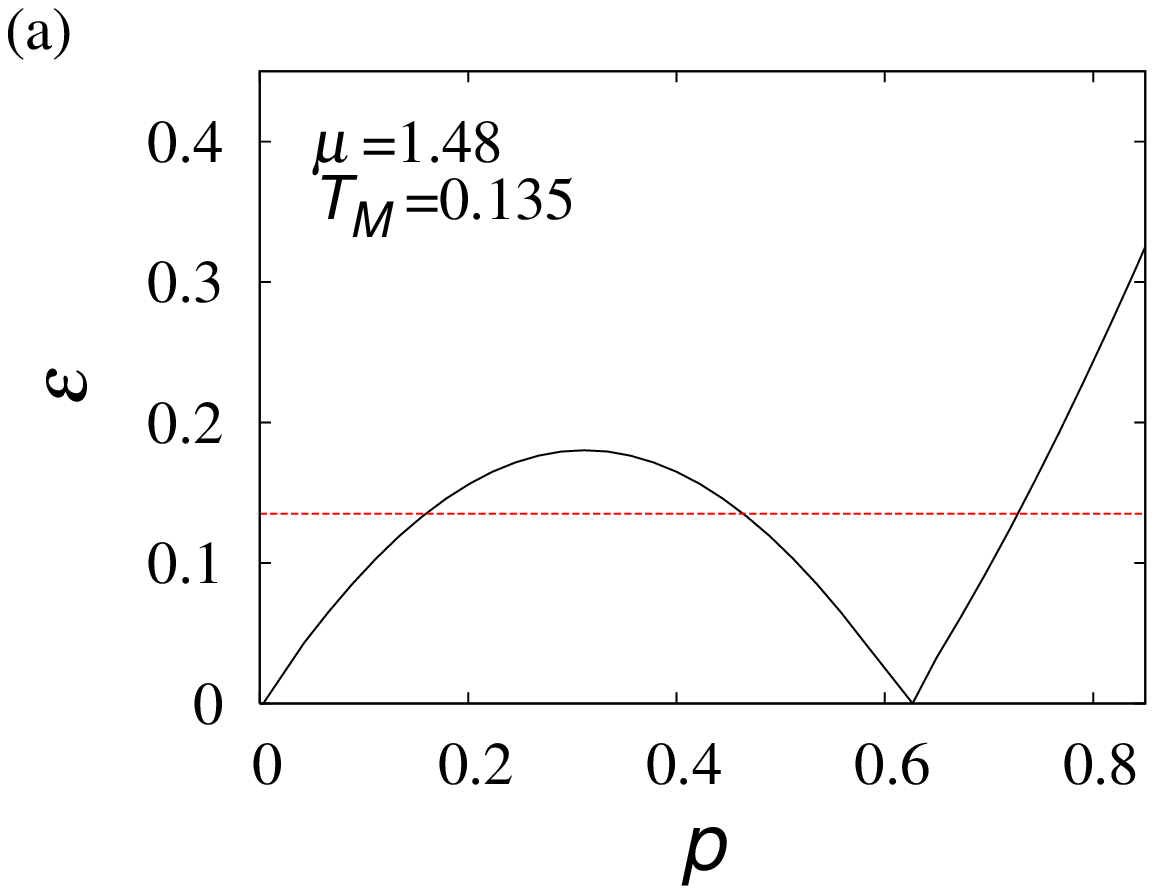}
\includegraphics[width=7.5cm]{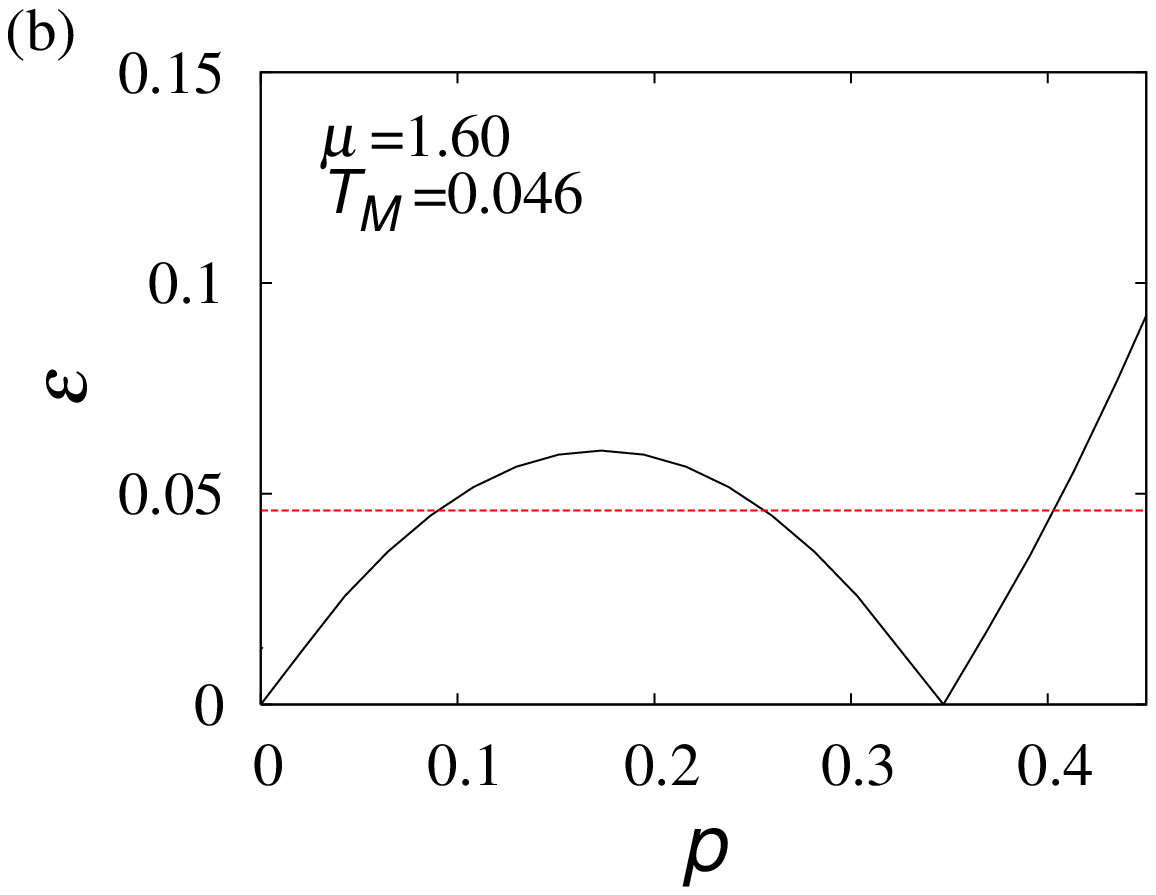}
\caption{(Color online) Lower edge of the charge excitation continuum for (a) $\mu=1.48$ and (b) $\mu=1.60$ in the 1D Hubbard model. The dotted lines are the excitation energies corresponding to $T_{\rm M}$. }
\label{sus}
\end{center}
\end{figure}
It is considered that the TLL is realized below $T_{\rm M}$. We investigate the TLL behavior from the viewpoint of the excitation spectra. 
The charge excitation spectra of the 1D Hubbard model are formulated using the Bethe ansatz solution \cite{book}. The excitation continuum can be obtained from the excitation energy and momentum: 
$\varepsilon(k_{\rm h}, k_{\rm p})=\kappa(k_{\rm p})-\kappa(k_{\rm h})$ and $p(k_{\rm p},k_{\rm h})=2\pi \int_{k_{\rm h}}^{k_{\rm p}} \rho(k) dk$, respectively, where $-Q \leq k_{\rm h} \leq Q$, $-\pi \leq k_{\rm p} \leq -Q$, and $Q \leq k_{\rm p} \leq \pi$. 
In Fig. \ref{sus}, we show the lower edge of the charge excitation continuum for two chemical potentials plotted in Fig. 2. 
The excitation energy ($\varepsilon_{\rm M}$) corresponding to $T_{\rm M}$ for a given chemical potential is also shown in the figure by the dotted line. 
We find that for $\varepsilon<\varepsilon_{\rm M}$, the lower edge of the charge excitation continuum nearly obeys the linear dispersion relation, which is a characteristic of the TLL, although the deviation may be pronounced close to $\varepsilon \sim \varepsilon_{\rm M}$. 
As shown in Appendix, also in the $S=1/2$ Heisenberg model with a large Ising anisotropy in magnetic fields, the lower edge of the excitation continuum for $H \gtrsim H_{\rm c}$ obeys the linear dispersion relation at $\varepsilon < \varepsilon_{\rm m}$, where $\varepsilon_{\rm m}$ is the excitation energy corresponding to the temperature $(T_{\rm m})$ for the magnetization minimum. 
In this way, the TLL for $T < T_{\rm m}, T_{\rm M}$ in 1D gapped many-body systems has been confirmed from the spectral point of view.

\section{\label{sec:level3}Summary}
We have investigated the temperature dependence of electron number in the 1D Hubbard model using the Bethe ansatz method. 
We have found a maximum structure at $\mu \lesssim \mu_{\rm c2}$. As chemical potential approaches $\mu_{\rm c2}$ from below, $T_{\rm M}$ approaches the universal relation asymptotically. 
We have also confirmed for the 1D Hubbard model and 1D $S=1/2$ Heisenberg model with a large Ising anisotropy that the linear dispersion relation is satisfied below the excitation energy corresponding to $T_{\rm m}, T_{\rm M}$. 
Judging from the findings obtained so far for 1D gapped many-body systems, the thermodynamic quantity corresponding to the particle number shows a minimum/maximum as a function of the temperature close to the critical point and the TLL is realized at $T \lesssim T_{\rm m}, T_{\rm M}$.

\section*{Acknowledgments}
We would like to thank S. Miyashita for helpful comments and valuable discussions. Numerical computations were carried out at the Supercomputer Center, the Institute for Solid State Physics, University of Tokyo. 
This work was partially supported by a Grant-in-Aid (No. 20540390) for Scientific Research from the Ministry of Education, Culture, Sports, Science, and Technology, Japan. 

\appendix
\section{Excitation spectra of the $S=1/2$ Heisenberg model with a large Ising anisotropy in magnetic fields}
\begin{figure}[htb]
\begin{center}
\includegraphics[width=7.5cm]{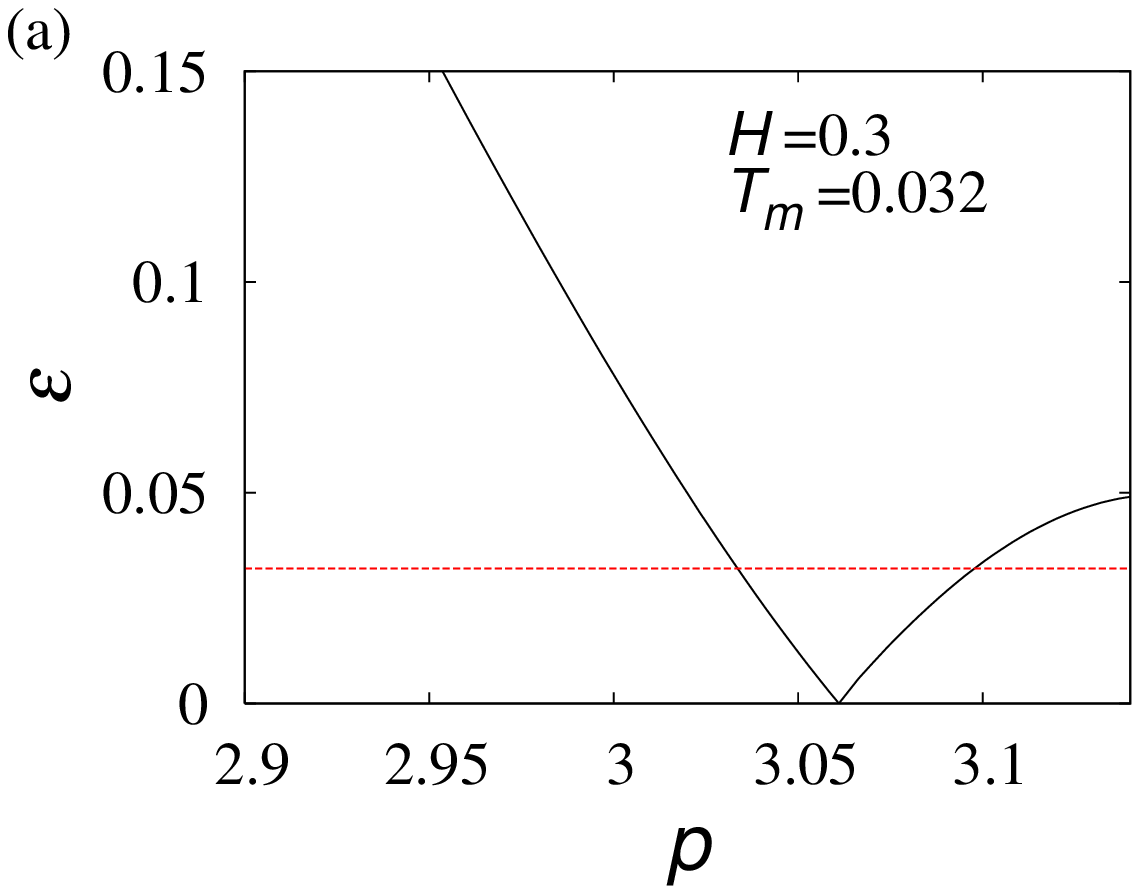}
\includegraphics[width=7.5cm]{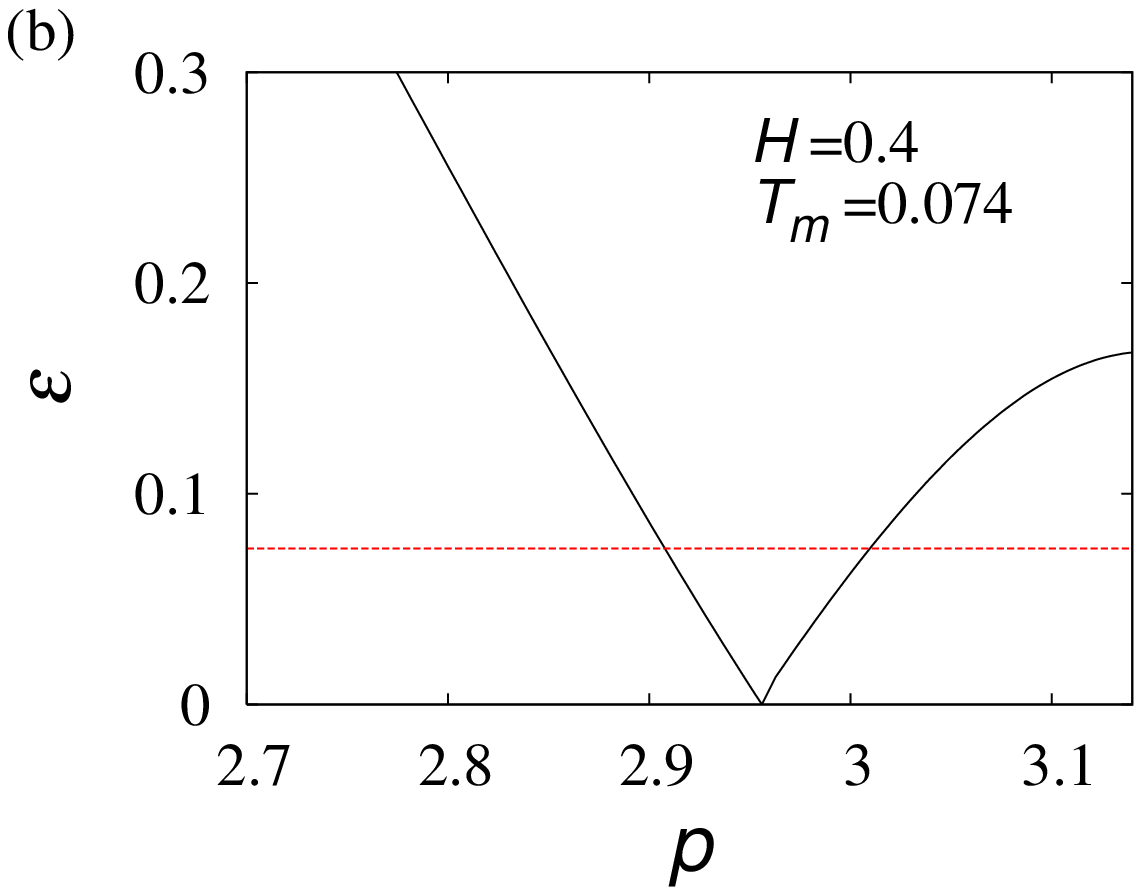}
\caption{(Color online) Lower edge of the excitation continuum for (a) $H=0.3$ and (b) $H=0.4$ in the $S=1/2$ Heisenberg model with a large Ising anisotropy: $\Delta=2.17$. The dotted lines are the excitation energies corresponding to $T_{\rm m}$. We use $g\mu_{\rm B}=2$. The antiferromagnetic exchange interaction is given in units of energy. }
\label{spc}
\end{center}
\end{figure}
We calculate the excitation spectra of the $S=1/2$ Heisenberg model with a large Ising anisotropy in magnetic fields using the Bethe ansatz solution. 
The Hamiltonian is given by 
\begin{eqnarray}
\mathcal{H} &=& J\sum_{i} \left(S^{x}_{i}S^{x}_{i+1}+S^{y}_{i}S^{y}_{i+1}
             + \Delta S^{z}_{i}S^{z}_{i+1} \right)   \nonumber \\
             &-& g\mu_{\rm B}\sum_{i}S^{z}_{i}, 
\end{eqnarray}
where $J>0$ and $\Delta=2.17$. We use $g\mu_{\rm B}=2$ and give $J$ in units of energy. 
Excitation energy and momentum are expressed using the dressed energy $\eta(\lambda)$ and the distribution function of the spin rapidity $\sigma(\lambda)$ as 
$\varepsilon(\lambda_{\rm h}, \lambda_{\rm p})=\eta(\lambda_{\rm p})-\eta(\lambda_{\rm h})$ and $p(\lambda_{\rm p},\lambda_{\rm h})=2\pi \int_{\lambda_{\rm h}}^{\lambda_{\rm p}} \sigma(\lambda) d\lambda$, respectively, where $-B \leq \lambda_{\rm h} \leq B$, $-\pi/\phi \leq \lambda_{\rm p} \leq -B$, and $B \leq \lambda_{\rm p} \leq \pi/\phi$ with $\phi=\cosh^{-1}\Delta$ $(\phi >0)$. 
The cutoff $B$ is obtained from the condition $\eta(B)=0$, where $\eta(\lambda)$ satisfies the integral equation 
$
\eta(\lambda)=\eta_0(\lambda)-\int_{-B}^{B} \, d\lambda^{\prime} b_2(\lambda-\lambda^{\prime}) \eta(\lambda^{\prime}) 
$
with $\eta_0(\lambda)=2H-(2\pi\sinh \phi/\phi)b_1(\lambda)$ and $b_n(\lambda)=\sinh(n\phi)/[\cosh(n\phi)-\cos \lambda]$. 
The distribution function is obtained from the integral equation 
$
\sigma(\lambda) = b_1(\lambda) - \int_{-B}^{B} \, d\lambda^{\prime} b_2(\lambda-\lambda^{\prime}) \sigma(\lambda^{\prime}). 
$
The lower edge of the excitation continuum and $\varepsilon_{\rm m}$ corresponding to $T_{\rm m}$ are shown in Fig. \ref{spc}. 
We also find that in this model, the lower edge of the excitation continuum obeys the linear dispersion relation at $\varepsilon < \varepsilon_{\rm m}$.


\end{document}